\documentclass[twocolumn]{aastex61}

\usepackage{natbib}

\usepackage{times}
\usepackage{epstopdf}
\usepackage{subfigure}
\usepackage{amsmath}

\renewcommand{\ion}[2]{#1\,{\textsc{\romannumeral#2}}}

\newcommand{\goesLong}{\textit{Geostationary Operational Environmental Satellites}}
\newcommand{\rhessi}{\textit{RHESSI}}

\newcommand{\aia}{\textit{AIA}}
\newcommand{\aiaLong}{\textit{Atmospheric Imaging Assembly}}

\newcommand{\goes}{\textit{GOES}}

\newcommand{\sdo}{\textit{SDO}}
\newcommand{\sdoLong}{\textit{Solar Dynamics Observatory}}

\makeatletter
\newcommand*{\rom}[1]{\expandafter\@slowromancap\romannumeral #1@}
\makeatother

\shorttitle{The evolution of GOES light curves}
\shortauthors{Reep \& Toriumi}

\begin{document}


\title{The direct relation between the duration of magnetic reconnection and the evolution of \goes\ light curves in solar flares}

\author[0000-0003-4739-1152]{Jeffrey W. Reep}
\affiliation{National Research Council Postdoctoral Fellow, Space Science Division, Naval Research Laboratory, Washington, DC 20375, USA; \href{mailto:jeffrey.reep.ctr@nrl.navy.mil}{jeffrey.reep.ctr@nrl.navy.mil}}

\author[0000-0002-1276-2403]{Shin Toriumi}
\affiliation{National Astronomical Observatory of Japan, 2-21-1 Osawa, Mitaka, Tokyo 181-8588, Japan;
 \href{mailto:shin.toriumi@nao.ac.jp}{shin.toriumi@nao.ac.jp}}


\begin{abstract}
\goes\ soft X-ray light curves are used to measure the timing and duration of solar flare emission.  The timing and duration of the magnetic reconnection and subsequent energy release which drives solar flares are unknown, though the light curves are presumably related.  It is therefore critical to understand the physics which connects the two: how does the time scale of reconnection produce an observed \goes\ light curve?  In this work, we model the formation and expansion of an arcade of loops with a hydrodynamic model, which we then use to synthesize \goes\ light curves.  We calculate the FWHM and the e-folding decay time of the light curves and compare them to the separation of the centroids of the two ribbons which the arcade spans, which is representative of the size scale of the loops.  We reproduce a linear relation between the two, as found observationally in previous work.  We show that this demonstrates a direct connection between the duration of energy release and the evolution of these light curves.  We also show that the cooling processes of individual loops comprising the flare arcade directly affect the measured time scales.  From the clear consistency between the observed and modeled linearity, we conclude that the primary factors that control the flare time scales are the duration of reconnection and the loop lengths.
\end{abstract}



\section{Introduction}
\label{sec:introduction}

The \goesLong\ (\goes), first launched in 1975 and most recently with \goes-16 in November 2016, are a series of satellites maintained by the National Oceanic and Atmospheric Administration (NOAA).  Since \goes-8 in 1986 and with previous NOAA satellites, they have carried two X-ray sensors (XRS) which continuously monitor the solar X-ray flux in the wavelength bands 1-8\,\AA\ ($\approx$ 1.5-12\,keV) and 0.5-4\,\AA\ ($\approx$ 3-24\,keV).  The emission in these bands is primarily due to dense plasma at temperatures exceeding 10\,MK, and therefore the light curves are used to track the timing and evolution of solar flares.   

Observational relationships between flare evolution and geometry are often discussed since they might be the key to accessing the physics behind flare eruptions \citep[see reviews by][]{priest2002,fletcher2011,shibata2011}.  In a recent paper, \citet{toriumi2017} measured the full-width-at-half-maximum (FWHM) and the e-folding decay time of 1-8\,\AA\ \goes\ light curves in 51 flares larger than M5 and correlated them with the distance between the centroids of the two flare ribbons $d_{\text{ribbon}}$, as measured in the 1600 \AA\ bandpass with the \aiaLong\ (\aia, \citealt{lemen2012}) onboard the \sdoLong\ (\sdo, \citealt{pesnell2012}).   The distance $d_{\text{ribbon}}$ was measured over the duration of each event, rather than at one single instant, as detailed in that paper (see Figure 1 of that work).  For both time scales, there was a strong linear correlation with the separation of the ribbon centroids.  They found similar correlations between the time scales and the magnetic flux as well as the ribbon area.

A few processes drive the thermal evolution of coronal loops.  Initially, after heating begins, the temperature rises sharply while the density is approximately unchanged, driving a strong conductive flux out of the corona.  As evaporation carries plasma into the loop and the density rises significantly, radiation comes to dominate the energy losses while conduction becomes significantly less efficient \citep{antiochos1980, cargill1995}.  An enthalpy flux of plasma flowing out of the corona begins to power the transition region at later times, which leads to a scaling law that relates the temperature and density in the cooling phase, $T \propto n^{\delta}$ \citep{bradshaw2005}, where $\delta$ relates the relative strength of radiation to enthalpy flux \citep{bradshaw2010}.  Once it falls below a critical temperature, the loop is unable to sustain the cooling, and undergoes ``catastrophic cooling'' where the temperature falls drastically while the density remains nearly constant \citep{reale2012,cargill2013}.  

Not much attention has been paid so far to the facts that the coronal flux system involved in a flare is composed of multiple individual, elementary loops, and that the loops start to get heated impulsively (probably due to magnetic reconnection) with a certain delay, as is obvious from the expansion of flare ribbons \citep{dodson1949,bruzek1964,grigis2005}.  One of the earliest attempts to model this was made by \citet{hori1997,hori1998}, who found that single loops were unable to explain observed blue-shifted \ion{Ca}{19} line profiles with a large stationary component, whereas a bundle of loops naturally does so.  \citet{reeves2007} employed a reconnection model to drive the heat input into a multi-threaded simulation in order to reproduce emissions seen by multiple instruments, and found that the reconnection model predicted the correct magnitude of energy release.   

In this work, we model the evolution of flare loops by introducing not only the thermal processes (radiation, conduction, and enthalpy drainage) but also the effect of successive reconnection, and synthesize the \goes\ light curves.  With these simulations, we test the relation directly: why is the separation of the ribbon centroids $d_{\text{ribbon}}$ correlated with the FWHM and e-folding decay time?  Assuming that the emission comes from many different loops of varying lengths, we synthesize a total \goes\ light curve, determine the time scales, and measure how they compare with $d_{\text{ribbon}}$.  We reproduce the linear correlations, and confirm that the loop lengths play an important role in determining the FWHM and decay time.  Since the duration of reconnection and subsequent expansion of the arcade determine the separation of the ribbon centroids, this relation demonstrates the direct connection between the duration of energy release and light curves in flares.

\section{Observations}
\label{sec:observations}

To further motivate this work, we briefly present \goes\ observations that can be later contrasted with the model that we have developed.  We employ two time scales to measure the rate of decay of the \goes\ SXR light curves.  The first, $\tau_{\text{FWHM}}$, is the standard full-width-at-half-maximum of the light curve, and the time $t_{\text{end}}$ past the peak with flux at half maximum is referred to as the ``\goes\ end time.''  The second time scale, $\tau_{\text{decay}}$, measures the e-folding decay time at the \goes\ end time:
\begin{equation}
\tau_{\text{decay}} = \left. - \frac{F_{\text{SXR}}(t)}{dF_{\text{SXR}}(t)/dt} \right\rvert_{t = t_{\text{end}}}
\end{equation}

The \goes\ flux is primarily due to thermal emission (bremsstrahlung and spectral lines) from dense, hot plasma, which is most sensitive to temperatures exceeding 10 MK (\citealt{garcia1994,white2005}; see Figure 1 of \citealt{warren2004} for a plot of the temperature dependence of each channel).  These two time scales therefore contain information about the rates at which the flaring loops heat to and cool from high temperatures.  The information is obscured, though, since many loops comprise the system at any given time, and because there is a finite non-thermal component that does not strongly depend on the plasma temperature.  The FWHM $\tau_{\text{FWHM}}$ measures both the rising and falling time of the light curve, and so is determined by the rate at which plasma heats and then cools.  The e-folding decay time $\tau_{\text{decay}}$ more directly measures the rate at which the light curve falls, and therefore is determined primarily by the cooling of the flaring loops.  

\begin{figure*}
\begin{minipage}[b]{0.333\linewidth}
\centering
\includegraphics[width=\textwidth]{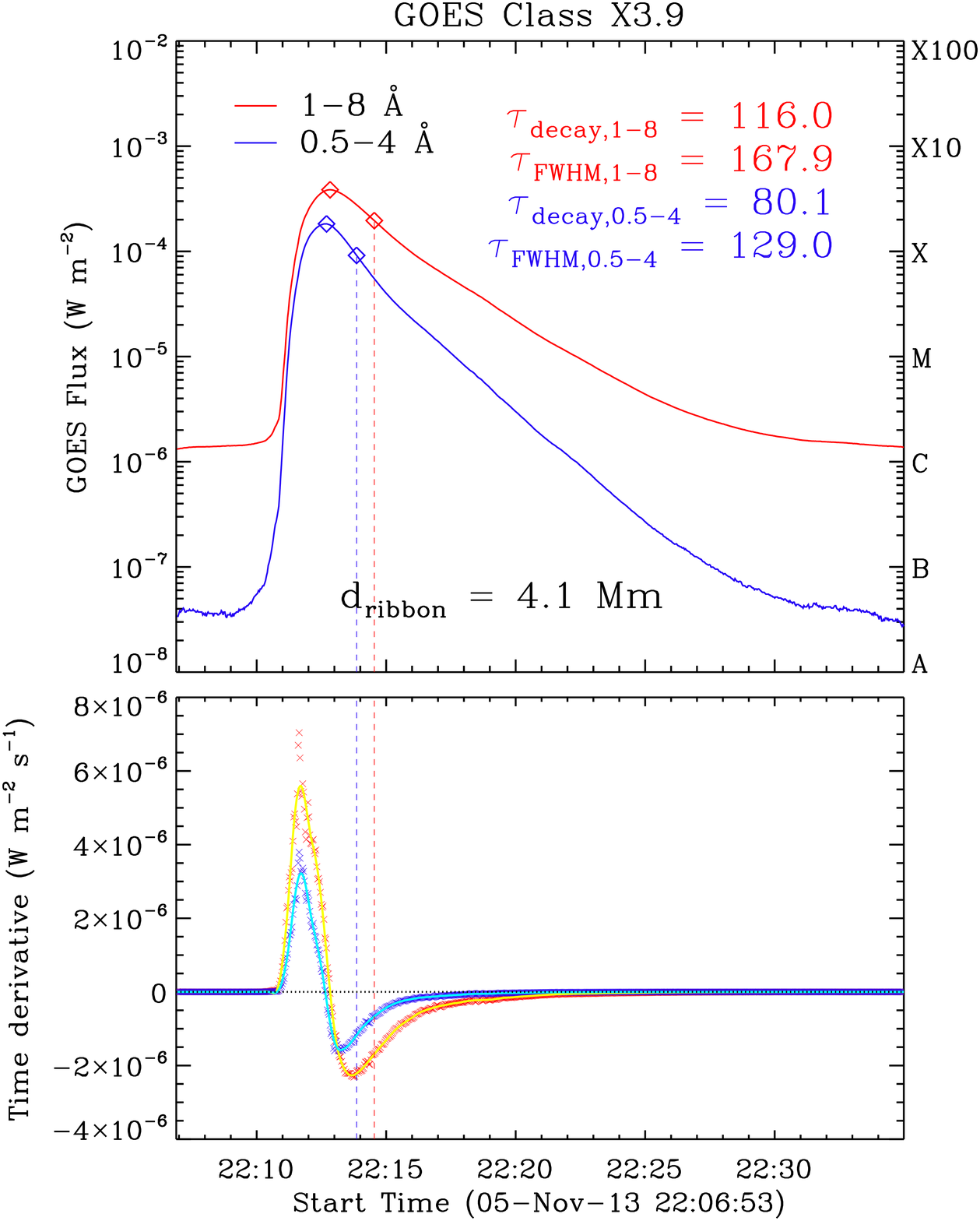}
\end{minipage}
\begin{minipage}[b]{0.333\linewidth}
\centering
\includegraphics[width=\textwidth]{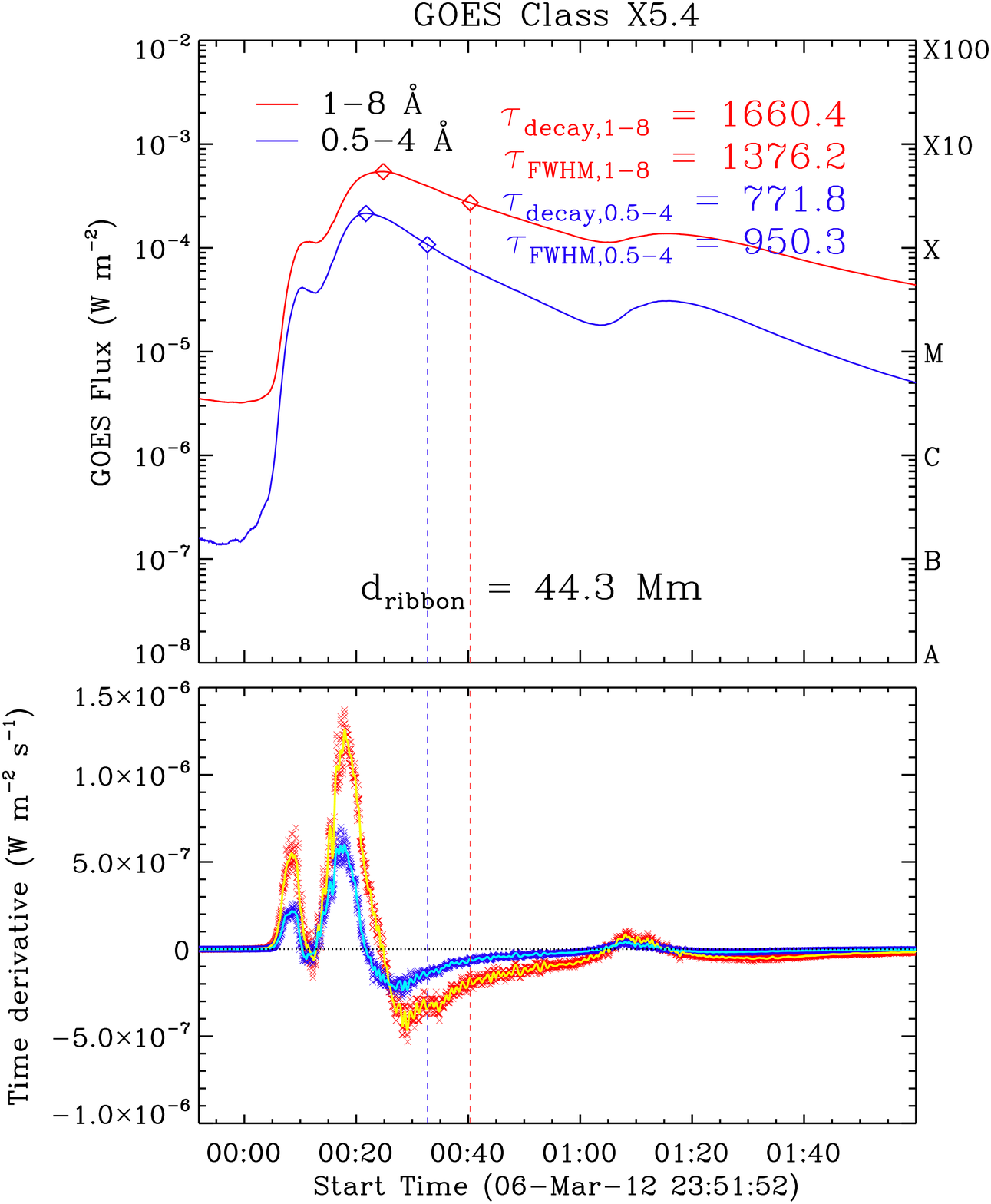}
\end{minipage}
\begin{minipage}[b]{0.333\linewidth}
\centering
\includegraphics[width=\textwidth]{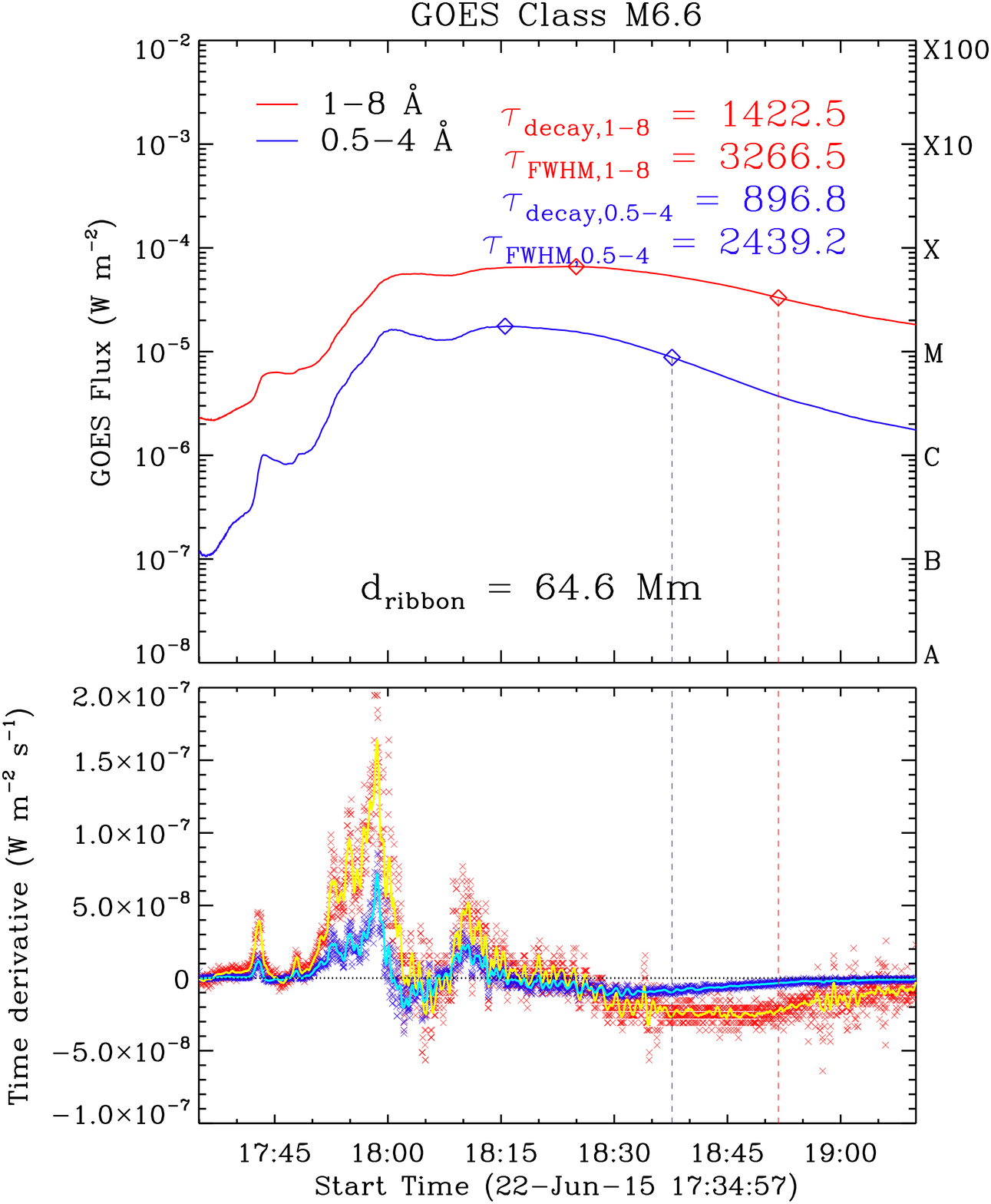}
\end{minipage}
\caption{\goes\ light curves (top) and time derivatives (bottom) for three large flares.  The 1-8\,\AA\ channel is shown in red, 0.5-4\,\AA\ channel in blue.  The calculated time scales, measured in seconds, are shown at the top right.  The derivatives have been calculated numerically (x marks), and with a Savitzky-Golay smoothing filter (yellow/cyan lines).  On the light curves, we have marked the times of peak emission and $t_{\text{end}}$ with diamonds.  We have drawn lines below $t_{\text{end}}$ to denote the times at which the derivative is used to calculate $\tau_{\text{decay}}$. }
\label{fig:obs_goes}
\end{figure*}

In Figure \ref{fig:obs_goes}, we show example \goes\ light curves for three flares: the X3.9 flare on 5 November 2013, the X5.4 on 6 March 2012, and the M6.6 on 22 June 2015.  The top plots show the light curves in each channel, 1-8\,\AA\ in red, 0.5-4\,\AA\ in blue, with the calculated time scales (measured in seconds) shown at top right.  The bottom plots show the time derivatives of each channel, calculated numerically (x marks), and with a Savitzky-Golay smoothing filter of degree 4 with 33 points \citep[yellow/cyan lines,][]{savitzky1964,press1986}.  We have also marked the times of peak emission and the \goes\ end times $t_{\text{end}}$ with diamonds.  Below $t_{\text{end}}$, we have drawn thin dashed lines to show the times at which the derivatives are used to calculate $\tau_{\text{decay}}$.  The total durations of these flares range from roughly 20 minutes to well over 3 hours, reflected in each respective FWHM.  For each case, we note the value of $d_{\text{ribbon}}$.

There are a few important features.  The \goes\ class has no bearing on either of the two time scales or the separation of the ribbon centroids $d_{\text{ribbon}}$.  The time scales in the 1-8\,\AA\ channel are systematically larger than those of the 0.5-4\,\AA\ channel, likely due to the differences in sensitivity to different temperatures, and because non-thermal emission is a relatively larger component of the high energy channel.  There is considerably more oscillatory behavior in the light curves for the longer duration events (particularly noticeable in the derivatives), which may be due to wave fluctuations \citep{mariska2006} or due to the variations of plasma temperature and density on individual threads \citep{rubiodacosta2016}.  These oscillations may be related to quasi-periodic pulsations \citep{nakariakov2009}, which have been reported in \goes\ observations \citep{dennis2017}.  Finally, there is generally a correlation between the magnitude of all of the time scales and $d_{\text{ribbon}}$, though there is considerable scatter.  

\begin{figure}
\begin{minipage}[b]{\linewidth}
\centering
\includegraphics[width=\textwidth]{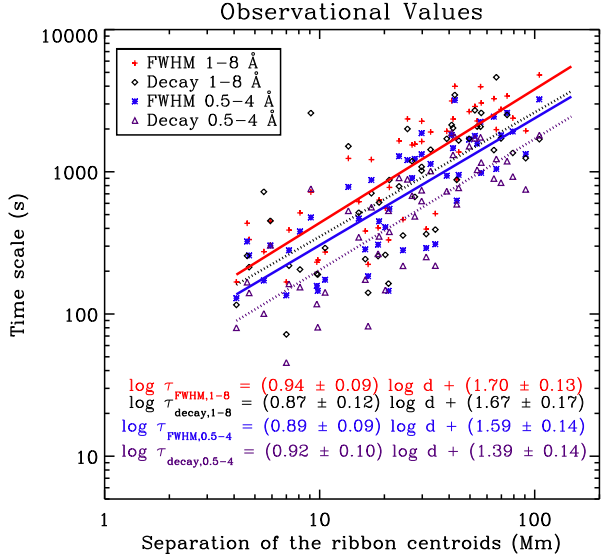}
\end{minipage}
\caption{The empirical results for 50 large flares.  We show the time scales for both \goes\ channels.  We have recalculated all time scales independently.    The red plus signs (blue asterisks) show the measured FWHM for the 1-8 (0.5-4) \AA\ channel, while the black diamonds (purple triangles) show the measured decay times.  The lines show the linear fits to the data.  The linear regression fits with 1-sigma uncertainties are shown for each time scale.  }
\label{fig:empirical}
\end{figure}

We present the empirical relationship in Figure \ref{fig:empirical} between the time scales of \goes\ 1-8 \AA\ light curves and separation of the ribbon centroids for 50 of the 51 flares studied by \citet{toriumi2017}, excluding event 12 which does not have a well-defined FWHM.  We use their values for $d_{\text{ribbon}}$, and independently have recalculated all time scales.  We have extended the analysis to include the high energy 0.5-4 \AA\ channel, and find similar results: there is an approximately linear correlation, though the magnitudes of the time scales are lower than those in the low energy channel.  The fits, also shown on the plot, are:
\begin{align}
\log{\tau_{\text{FWHM, 1-8 \AA}}} &= (0.94 \pm 0.09) \log{d_{\text{ribbon}}} + (1.70 \pm 0.13) \nonumber \\
\log{\tau_{\text{decay, 1-8 \AA}}} &= (0.87 \pm 0.12) \log{d_{\text{ribbon}}} + (1.67 \pm 0.17) \nonumber \\
\log{\tau_{\text{FWHM, 0.5-4 \AA}}} &= (0.89 \pm 0.09) \log{d_{\text{ribbon}}} + (1.59 \pm 0.14) \nonumber \\
\log{\tau_{\text{decay, 0.5-4 \AA}}} &= (0.92 \pm 0.10) \log{d_{\text{ribbon}}} + (1.39 \pm 0.14) 
\end{align} 
\noindent The trend is approximately linear in all four cases, though it is unclear why.  We now turn our focus towards explaining this result: how does this correlation arise?

\section{Ribbon Expansion Model}
\label{sec:modeling}

\begin{figure}
\begin{minipage}[b]{\linewidth}
\centering
\includegraphics[width=0.9\textwidth]{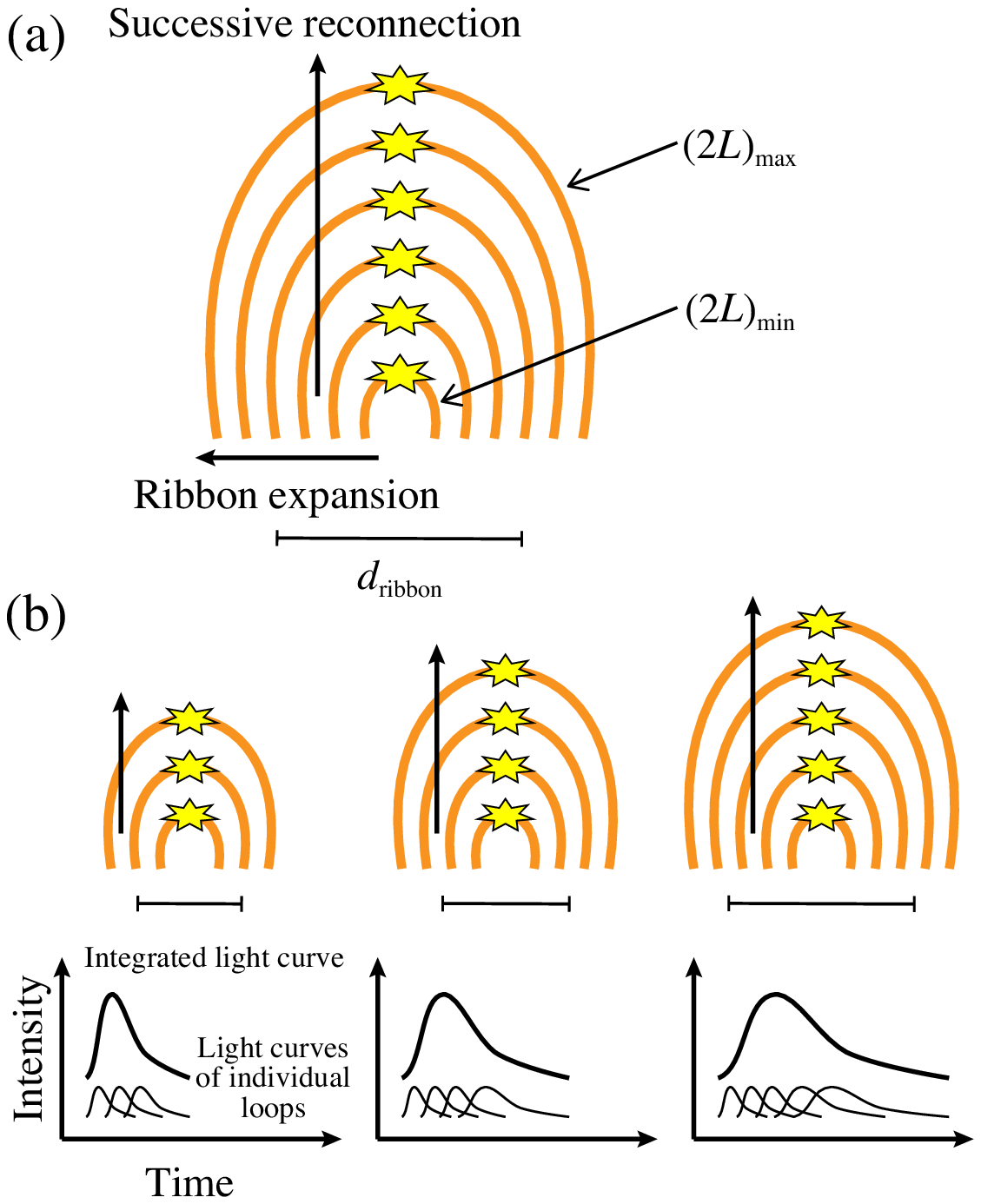}
\end{minipage}
\caption{A schematic cartoon illustrating the method by which we create bundles of loops that are used to synthesize the integrated \goes\ light curves. (a) As the reconnection event proceeds and the ribbons expand, new loops of increasing lengths form and are energized successively. (b) The intensity from each light curve contributes to the total observed light curve of the flare, which is then used to calculate the two time scales. We synthesize the light curves for many bundles with various separations between the ribbon centroids $d_{\text{ribbon}}$.}
\label{fig:cartoon}
\end{figure}

In order to test the relation between ribbon separation and the time scales, we synthesize \goes\ light curves for bundles of loops with varying separations between the ribbon centroids $d_{\text{ribbon}}$.  We first synthesize light curves of individual loops, which we then use to construct a composite light curve assuming a bundle of loops with ever-increasing lengths as the reconnection event proceeds.  Figure \ref{fig:cartoon} illustrates the scheme we use to construct the bundles and integrated light curves.    

The separation between centroids $d_{\text{ribbon}}$ for a bundle can be found from simple geometry since we assume that the loops are all semi-circular.  We take the centroid of a bundle to be the the foot-point separation of a loop with the median length of the loops in the bundle.  In other words, $d_{\text{ribbon}}$ is the diameter of a loop with length $\frac{(2L)_{\text{max}} + (2L)_{\text{min}}}{2}$, so it is straight-forward to find that $d_{\text{ribbon}} = \frac{(2L)_{\text{max}} + (2L)_{\text{min}} }{\pi}$.

We define a reconnection time scale $\tau_{\text{rec}}$ as 
\begin{equation}
\tau_{\text{rec}} = \frac{\zeta_{\text{ribbon}}}{V_{\text{ribbon}}} = \frac{\frac{1}{\pi}\Big((2L)_{\text{max}} - (2L)_{\text{min}}\Big)}{V_{\text{ribbon}}}
\end{equation}
\noindent which simply says that the time to expand outwards is determined by the distance $\zeta_{\text{ribbon}} = \frac{1}{\pi}\Big((2L)_{\text{max}} - (2L)_{\text{min}}\Big)$ from the initial reconnection site to the maximum extent of the ribbon, as well as by the ribbon expansion speed $V_{\text{ribbon}}$.  Assuming that the reconnection starts from a low altitude, \textit{i.e.} in the limit $L_{\text{min}} \rightarrow 0$, we have 
\begin{equation}
\tau_{\text{rec}} = \frac{\frac{1}{\pi} (2L)_{\text{max}}}{V_{\text{ribbon}}} \approx \frac{d_{\text{ribbon}}}{V_{\text{ribbon}}}
\end{equation}
\noindent or that $\tau_\text{rec} \propto {L_{\text{max}}} \propto d_{\text{ribbon}}$, which demonstrates the relation between the duration of reconnection and $d_{\text{ribbon}}$.  Assuming a constant ribbon speed of $V_{\text{ribbon}} = 20$\,km\,s$^{-1} = 0.02$\,Mm\,s$^{-1}$, we have 
\begin{equation}
\tau_{rec} \approx 50 \times d_{\text{ribbon}} 
\end{equation}
\noindent with distances measured in Mm and time scales in s.  Observationally, we have found 
\begin{align}
\log{\tau_{\text{FWHM, 1-8\AA}}} &= 0.94 \times \log{d_{\text{ribbon}}} + 1.70  \\ \nonumber
\rightarrow \tau_{\text{FWHM, 1-8\AA}} &= 10^{1.70} \times (d_{\text{ribbon}})^{0.94} \\ \nonumber
 &\approx 50 \times d_{\text{ribbon}}
\end{align}
\noindent  The striking agreement between the ribbon expansion scenario and the observational results points to the possibility that the duration of reconnection is one of most important parameters controlling the flare time scales.  We therefore test this directly with hydrodynamic simulations, from which we synthesize the SXR emission as might be seen by \goes.

\subsection{Hydrodynamic modeling}
\label{subsec:HD}

To test the cause of the correlations, we have run numerical experiments with the HYDrodynamics and RADiation code (HYDRAD, \citealt{bradshaw2003}).  The code solves the equations of conservation for mass, momentum, electron and ion energy along a one-dimensional magnetic flux tube \citep{bradshaw2013}.  HYDRAD employs adaptive mesh refinement to properly resolve the transition region (TR), which is vitally important for determining accurate coronal densities \citep{bradshaw2013}.  The code accounts for radiative losses calculated with CHIANTI version 8 \citep{dere1997,delzanna2015}, which includes full non-equilibrium ionization of various ion species, which is important for cooling rates \citep{bradshaw2004, bradshaw2013b}.  In this work, we assume all loops are semi-circular, oriented vertically relative to the solar surface.    

We assume that the primary heating mechanism is a beam of non-thermal electrons that deposit their energy through Coulomb collisions in the upper chromosphere \citep{emslie1978}.  We have assumed an injected electron spectrum of the form (electrons\,s$^{-1}$\,cm$^{-2}$\,keV$^{-1}$):
\begin{equation}
\mathfrak{F}(E_{0}, t) = \frac{F_{0}(t)}{E_{c}^{2}}\ (\delta - 2) \times
  \begin{cases}
   0 & \text{if } E_{0} < E_{c} \\
   \Big(\frac{E_{0}}{E_{c}}\Big)^{- \delta}       & \text{if } E_{0} \geq E_{c}
  \end{cases}
\label{sharpdist}
\end{equation}
where $F_{0}(t)$ is the initial energy flux of the beam, $E_{c}$ is the low-energy cut-off, $\delta$ is the spectral index.  

We have run simulations of loops with coronal lengths $2L = $[3, 4, 5, 6, 7, 8, 9, 10, 15, 20, 25, 30, 35, 40, 50, 75, 100, 125, 150, 175, 200]\,Mm (and a chromospheric depth of 2.2 Mm).  We heat each loop with the same energy flux carried by the electron beam, $F_{0} = 10^{11}$\,erg\,s$^{-1}$\,cm$^{-2}$, and a spectral index $\delta = 4$.  We have run three sets of simulations, one with a low energy cut-off $E_{c} = 20$\,keV and 30 seconds of total heating, one with $E_{c} = 10$\,keV for 30 seconds of heating, and one with $E_{c} = 20$\,keV for 60 seconds of heating.  The cooling of individual loops depends only weakly on the heating rate ($\propto Q^{-1/6}$, \citealt{cargill1994}), so the assumption of constant energy flux should not drastically affect the measured time scales.  We also assume that the cross-sectional area of the loops grows with time (equivalently, that the ribbon area grows with time), reaching its maximum value at the time of peak SXR emission (see Fig. 2 of \citealt{toriumi2017}).  

We calculate the X-ray emissions as a sum of thermal and non-thermal components for a given loop as follows.  The thermal components are evaluated with CHIANTI version 8 \citep{dere1997,delzanna2015}, using the ``isothermal'' routine to generate tables of spectra as a function of temperature and emission measure, which includes contributions from continuum processes and line emission.  In each grid cell, we calculate the local emission measure $EM = n^{2} A \Delta s$, where $n$ is the density, $A$ the assumed cross-sectional area, and $\Delta s$ the width of the grid cell.  We then use the local emission measure and temperature to calculate the spectrum of that grid cell, which we sum across all the grid cells and at each time step to create a composite spectrum since \goes\ XRS is not spatially resolved.   

We then add the non-thermal emission to the spectra assuming a thick-target \citep{brown1971}.  Free-free emission can be written \citep{kontar2011,reep2013}
\begin{equation}
I_{\text{thick}} = \frac{A}{4\pi R^{2}} \int_{\epsilon}^{\infty} \frac{n_{H}\,v\,Q(\epsilon, E)\ dE}{dE/dt} \int_{E}^{\infty} \mathfrak{F}(E_{0}, t) dE_{0}
\end{equation}
where $R = 1$\,AU\ $ = 1.497 \times 10^{13}$\,cm, $\epsilon$ is the photon energy (keV), $n_{H}$ the hydrogen density (cm$^{-3}$), $v$ the electron velocity (cm\,s$^{-1}$), $Q(\epsilon, E)$ the cross-section of the interaction (cm$^{2}$\,keV$^{-1}$), and $\frac{dE}{dt}$ the energy lost by the electron per unit time (keV\,s$^{-1}$).  The inner integral is analytic for the assumed form of the injected electron spectrum.  For the outer integral, we use a fully relativistic Bethe-Heitler cross-section with the Elwert correction factor \citep{bethe1934,elwert1939,koch1959}, which is then evaluated using Gauss-Laguerre quadrature of at least 100th order.  

Although expected to be small, we include non-thermal free-bound emission using the formulation of \citet{brown2008,brown2009}, with corrections given by \citet{brown2010} and \citet{reep2016}.  We assume only recombination onto iron ions of ionization stage above \ion{Fe}{20}, which are the most significant contribution to this emission process \citep{reep2016}.  

Once all the emission has been summed, we then fold the spectra through the \goes\ response at the appropriate energy ranges to calculate light curves.  The observed flares were all measured with \goes-15, so we use the response functions for that satellite \citep{white2005}.

In order to create composite light curves, we must add together the emission from many loops.  We follow the basic scenario outlined in Figure \ref{fig:cartoon}.  Using a typical ribbon expansion speed of $\approx$ [10, 20, 30]\,km\,s$^{-1}$ (\textit{e.g.} \citealt{asai2004}), we assume that each new loop reconnects and is energized with a given time lag past the previous one.  We use a sub-set of the simulations to form the bundles, \textit{e.g.} with lengths [3, 4, 5]\,Mm, [3, 4, 5, 6]\,Mm, [3, 4, 5, 6, 7]\,Mm, \textit{etc.}.  

To reduce the number of required simulations, and to ensure reasonably smooth light curves, at each time we linearly interpolate the X-ray emission between these simulations as if a new loop were to form at regular intervals of $\frac{1}{3}$\,Mm.  For example, between the loops with $2L = 3$ and $4$\,Mm, we have 2 interpolated loops of lengths $3.33$ and $3.66$\,Mm.  For the first loop, we calculate its X-ray intensity $I_{\text{3.33\,Mm}}(t) = \frac{2}{3} I_{\text{3\,Mm}}(t) + \frac{1}{3} I_{\text{4\,Mm}}(t)$, and for the second, we similarly have $I_{\text{3.66\,Mm}}(t) = \frac{1}{3} I_{\text{3\,Mm}}(t) + \frac{2}{3} I_{\text{4\,Mm}}(t)$.  We similarly interpolate between each successive simulation in the bundle.

For all bundles, using the composite light curves, we calculate the FWHM and decay time for both \goes\ channels.  We then plot these values against the separation distance of the ribbon centroids $d_{\text{ribbon}}$ for each bundle, and fit a linear regression to measure any correlation.  Finally, we compare the measured correlations against the observations in Section \ref{sec:observations}.

\section{Results}
\label{sec:results}

\subsection{Synthesized light curves}
\label{subsec:light_curves}

\begin{figure*}
\begin{minipage}[b]{0.333\linewidth}
\centering
\includegraphics[width=\textwidth]{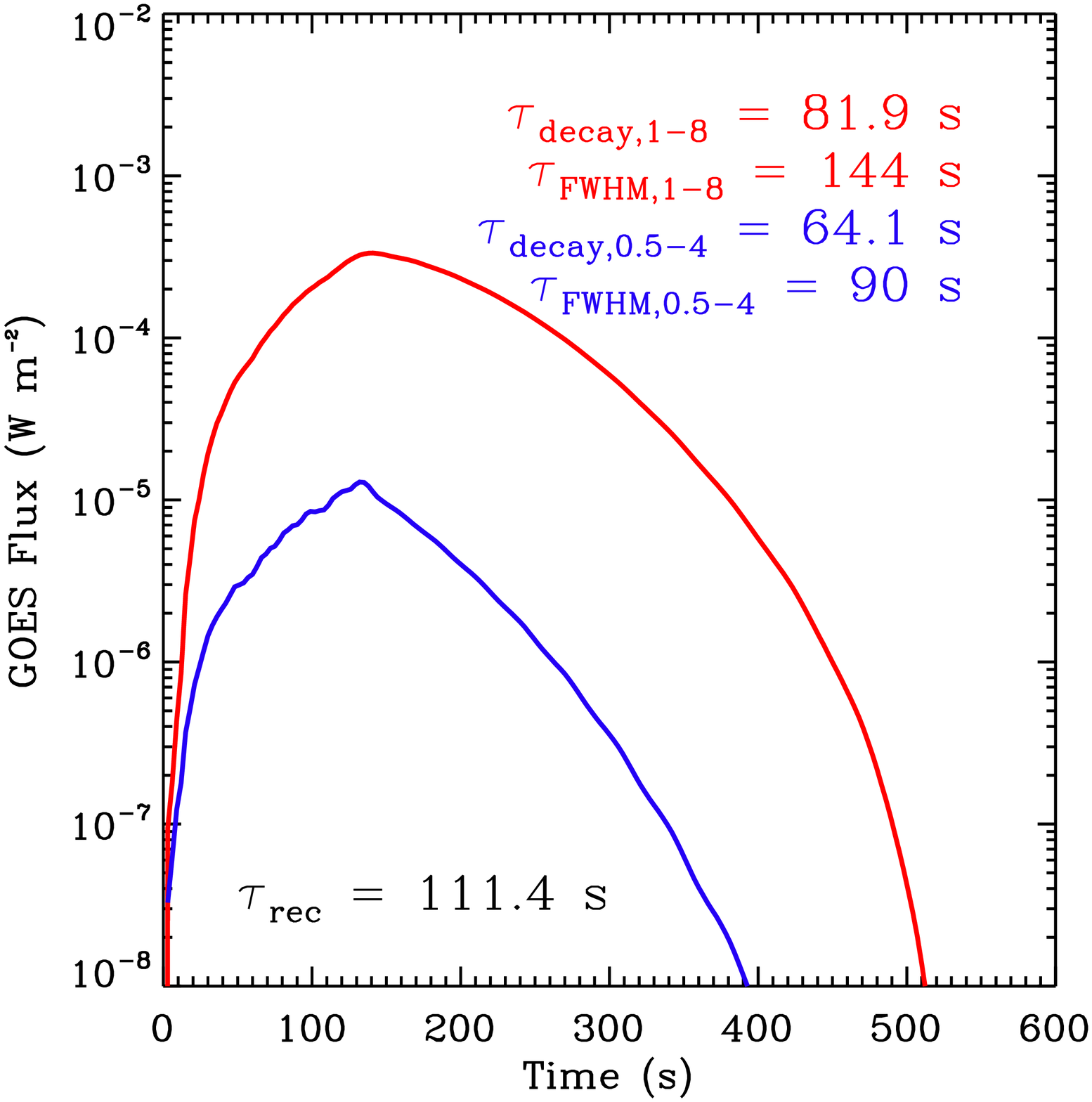}
\end{minipage}
\begin{minipage}[b]{0.333\linewidth}
\centering
\includegraphics[width=\textwidth]{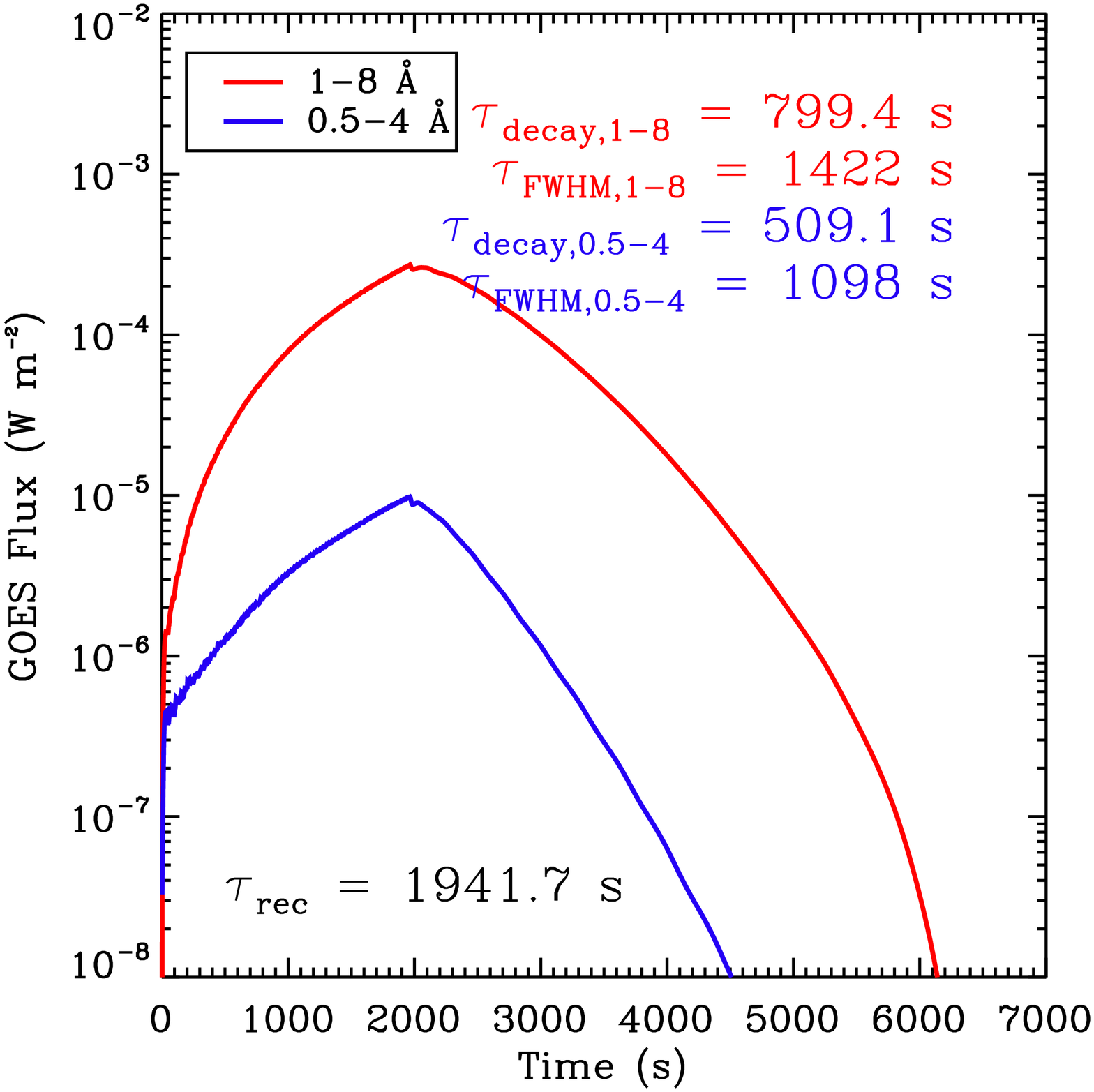}
\end{minipage}
\begin{minipage}[b]{0.333\linewidth}
\centering
\includegraphics[width=\textwidth]{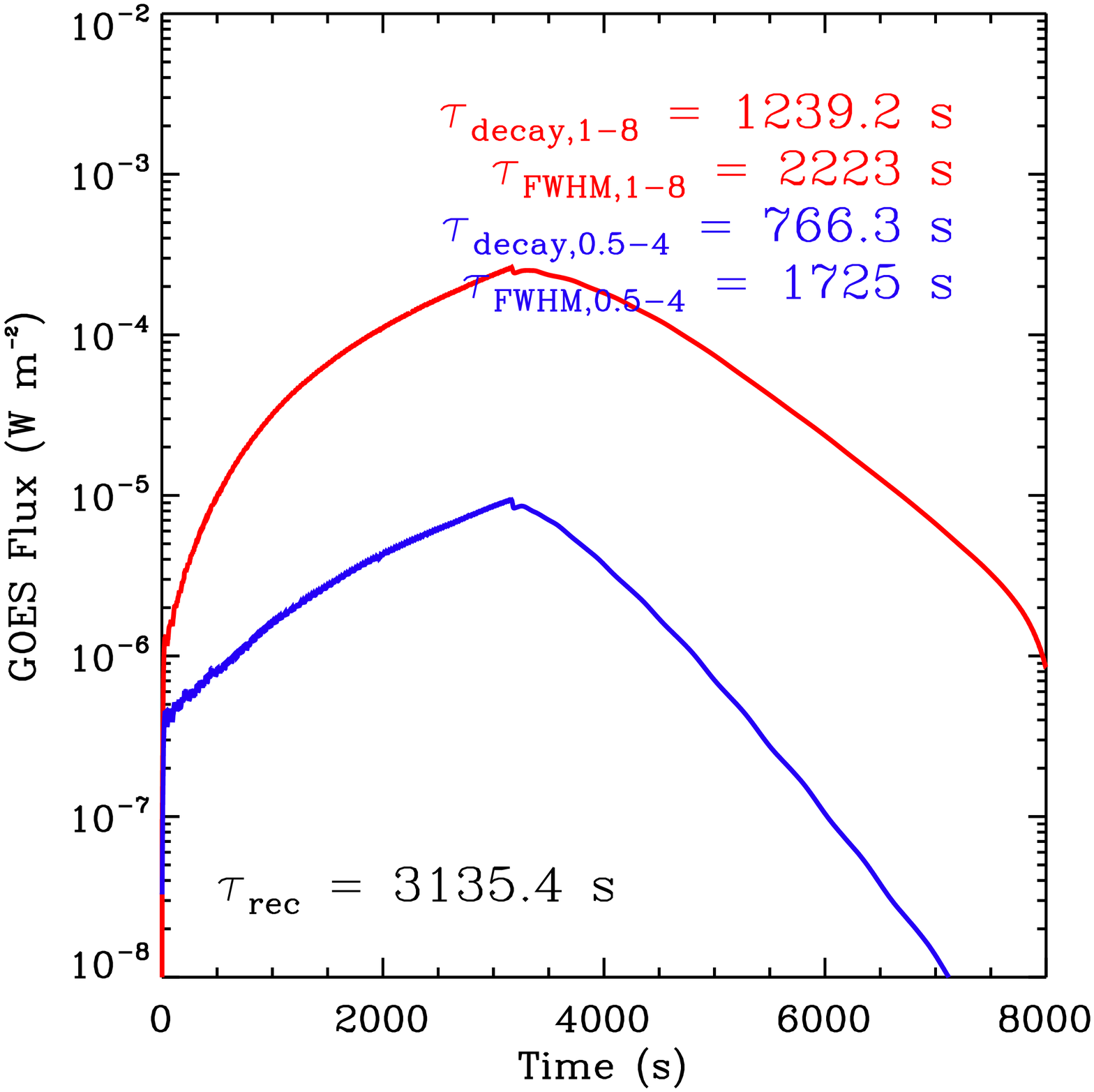}
\end{minipage}
\begin{minipage}[b]{0.333\linewidth}
\centering
\includegraphics[width=\textwidth]{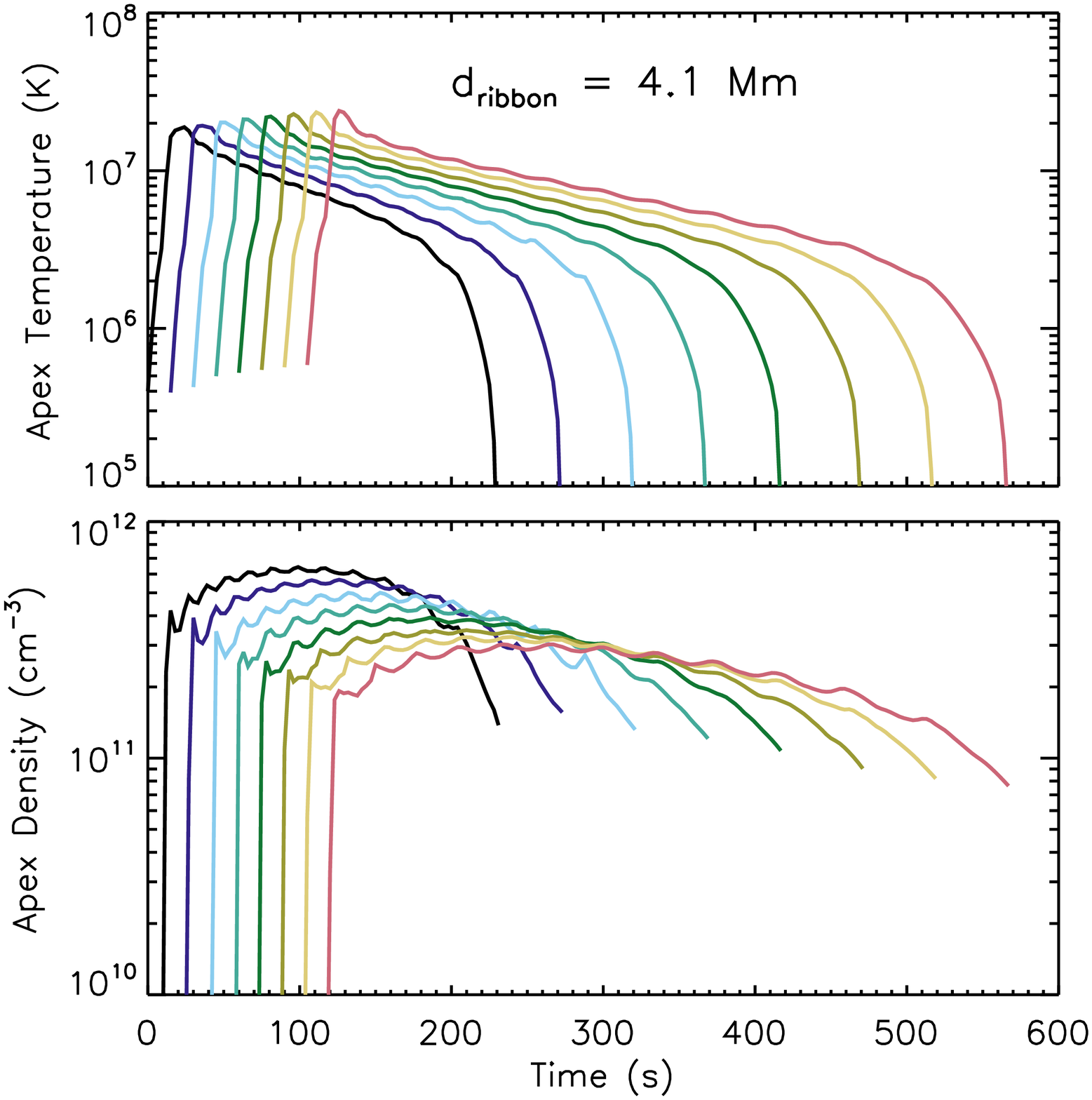}
\end{minipage}
\begin{minipage}[b]{0.333\linewidth}
\centering
\includegraphics[width=\textwidth]{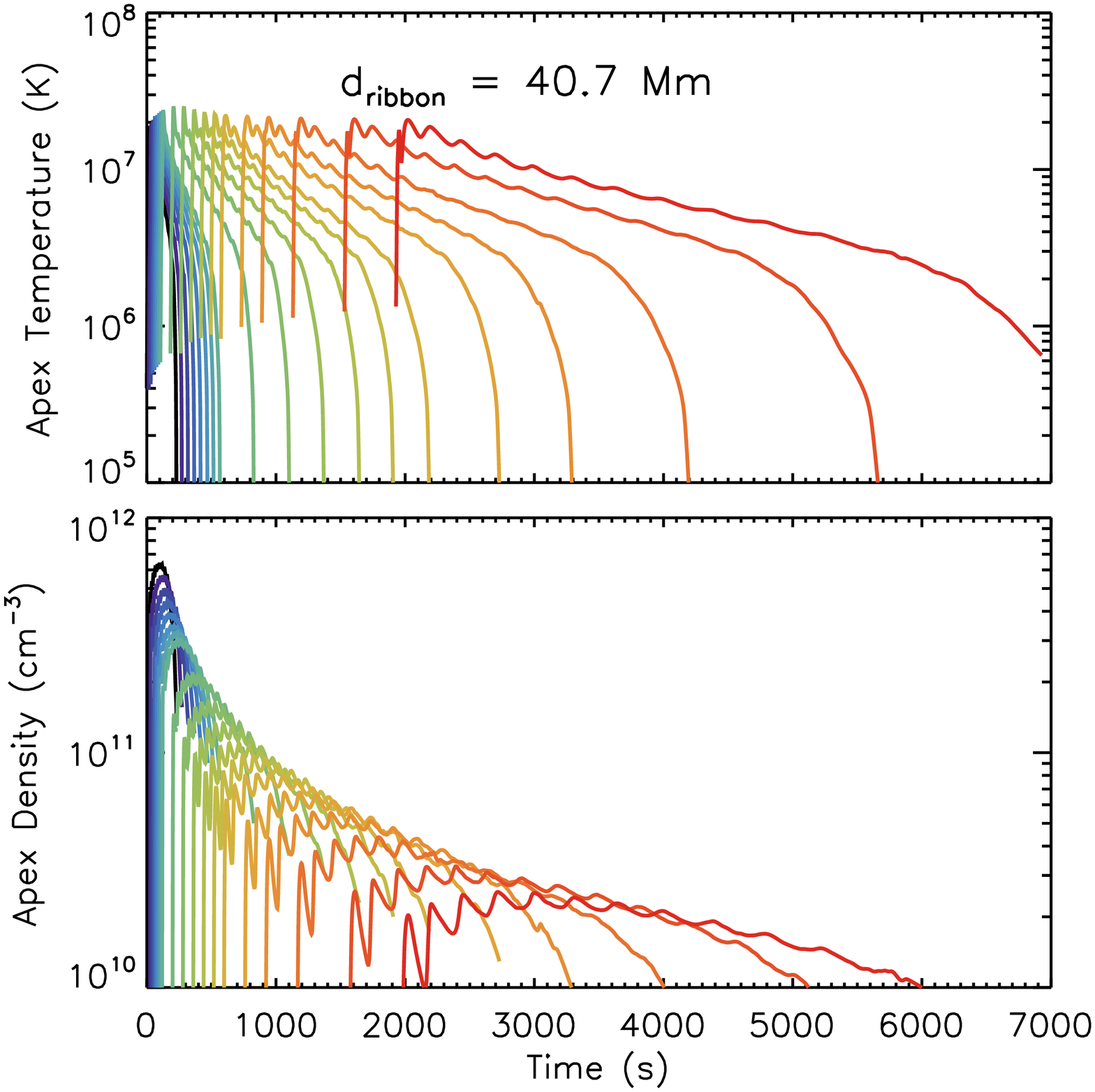}
\end{minipage}
\begin{minipage}[b]{0.333\linewidth}
\centering
\includegraphics[width=\textwidth]{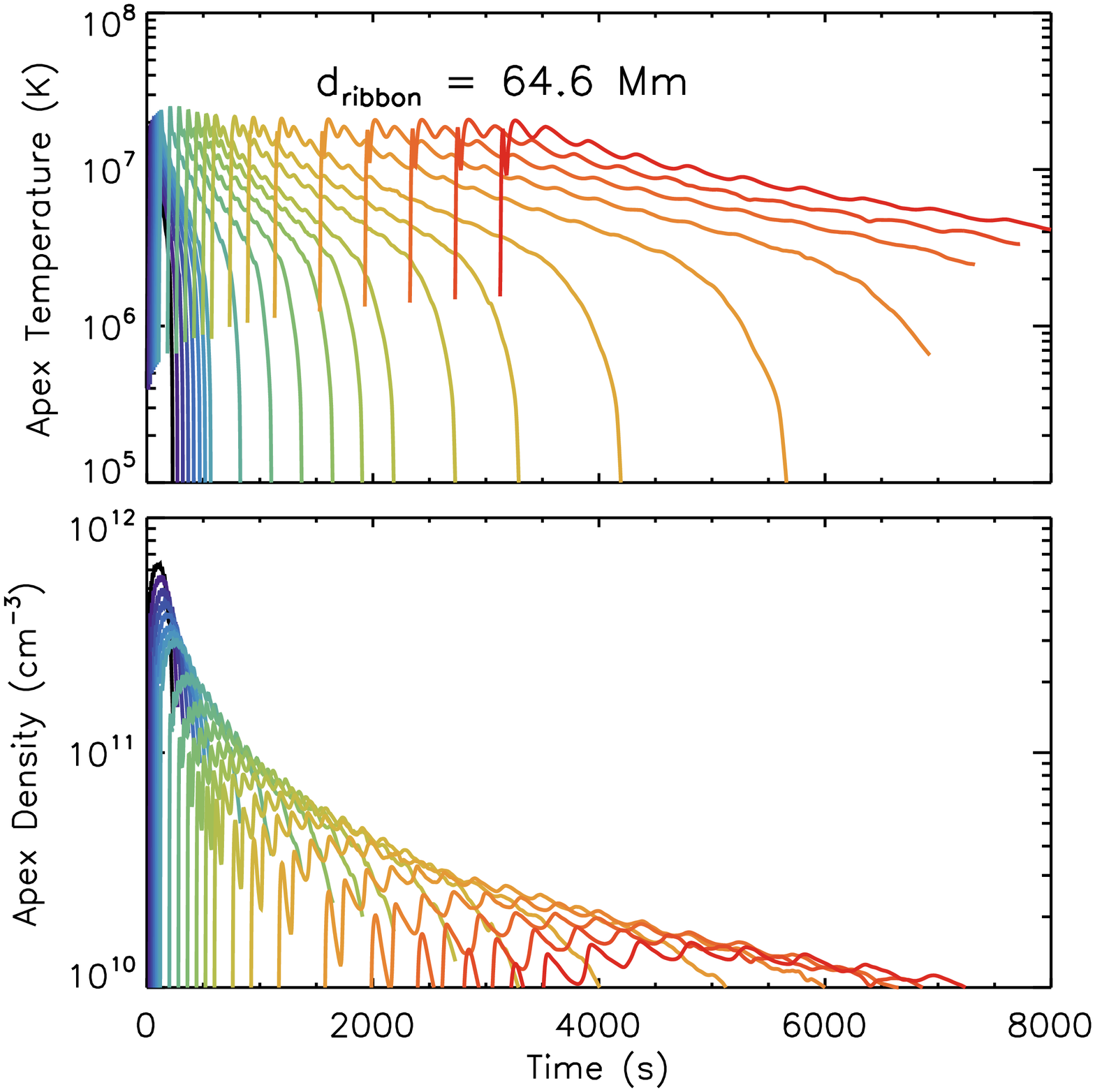}
\end{minipage}
\caption{At top, synthesized \goes\ light curves for three sets of loop bundles.  Each bundle has a minimum length $(2L)_{\text{min}} = 3$\,Mm, and maximum lengths of $(2L)_{\text{max}} = $[10, 125, 200]\,Mm, corresponding to $d_{\text{ribbon}} = $[4.1, 40.7, 64.6]\,Mm.  These assume a ribbon speed $V_{\text{ribbon}} = 20$\,km\,s$^{-1}$.  At bottom, the apex electron temperatures and densities in each loop comprising the arcade as a function of time. }
\label{fig:bundle_lightcurve}
\end{figure*}

We begin by showing a few example synthesized \goes\ light curves, along with the plasma evolution of the emitting threads.  Figure \ref{fig:bundle_lightcurve} shows three bundles of loops, with minimum length $(2L)_{\text{min}} = 3$\,Mm, and (respectively) maximum lengths $(2L)_{\text{max}} = $[10, 125, 200]\,Mm, corresponding to $d_{\text{ribbon}} = $[4.1, 40.7, 64.6]\,Mm.  The top row shows the synthesized \goes\ light curves in 1-8\,\AA\ (red) and 0.5-4\,\AA\ (blue), along with the measured FWHM $\tau_{\text{FWHM}}$ and e-folding decay time $\tau_{\text{decay}}$ for both channels (in seconds).  The bottom row shows the apex electron temperature and electron number density in each individual thread as a function of time (excluding the interpolated loops).  These light curves were synthesized assuming a ribbon speed of $V_{\text{ribbon}} = 20$\,km\,s$^{-1}$.

The evolution of the plasma follows a similar path for any individual loop.  Shortly after the onset of heating, the temperature rises sharply to over 20\,MK, causing the thermal emissions (primarily bremsstrahlung) to begin to rise, though the emission measure (EM) is initially small.  As the non-thermal electrons deposit their energy in the chromosphere, the pressure there expands as the temperature rises, which causes an expansion of material back into the corona, thus raising the EM and significantly increasing the thermal emissions.  As the coronal density rises, thermal conduction becomes significantly less effective, and radiative losses begin to dominate.  The density remains approximately constant while the temperature slowly falls during this time, which causes a reduction in the thermal emissions.  Eventually, the loop begins to catastrophically collapse, and the temperature plummets to chromospheric temperatures in a short time, so that the loop effectively disappears in SXRs.  

The total \goes\ emissions are spatially unresolved, so the light curves follow the evolution of many individual loops at any given time.  Both \goes\ channels are sensitive to high temperature emission (with significant overlap), but the 0.5-4\,\AA\ channel's sensitivity falls off more rapidly at lower temperatures (see Fig. 6 of \citealt{white2005} and Fig. 1 of \citealt{warren2004}).  Therefore, as the temperature decreases on the hottest loop, the light curve in 0.5-4\,\AA\ falls off more rapidly than in the 1-8\,\AA\ channel.  This in turn causes both the FWHM and decay time to be lower in the 0.5-4\,\AA\ channel (which also remains true in the observed flares).  

All four time scales are higher for larger centroid separations $d_{\text{ribbon}}$, though the shapes of the light curves are similar.  After the reconnection ceases and new loops are no longer being heated, plasma above $10$\,MK becomes scarce, significantly reducing the thermal bremsstrahlung which dominates the emission in the two channels.  The duration of reconnection, therefore, appears critical in determining the evolution of the \goes\ light curves.  

There is no late phase heating in these simulations, so the cooling proceeds rapidly once heating ceases, which causes the sharp decrease in intensity.  There are, however, indications that late phase heating occurs in flares, both in the TR \citep{doschek1977} and corona \citep{petrasso1979}.  This is clear from the sustained high temperatures \citep{svestka1982}, high densities \citep{moore1980}, and evaporative up-flows \citep{czaykowska1999,czaykowska2001}.  Modeling efforts were able to quantify the energy release, and show that the heating is consistent with sustained reconnection \citep{cargill1982,cargill1983,pneuman1982}.  There have recently been efforts to include late phase heating into multi-threaded modeling of flares \citep{qiu2016}, though the magnitude, temporal envelope, and duration of that heating require further study, a point which we plan to address in future work.

There is another important factor: the time scales also strongly depend on the lengths of the loops within a bundle.  To show this, we examine bundles of loops with an equal reconnection time scale, but variable $d_{\text{ribbon}}$.  $\tau_{\text{rec}}$ measures the duration of time between the formation of the first and last loop in the bundle, and therefore depends on the distance between them and the speed at which the ribbon spreads.  

\begin{figure}
\begin{minipage}[b]{\linewidth}
\centering
\includegraphics[width=\textwidth]{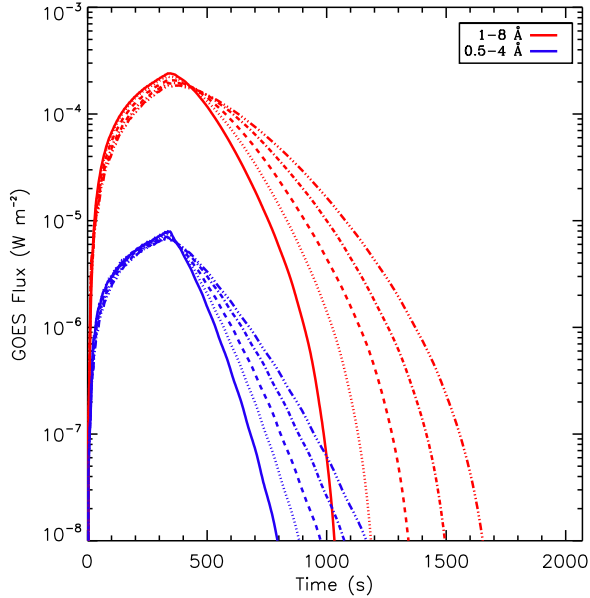}
\end{minipage}
\caption{Five bundles of loops with an equal reconnection time scale $\tau_{\text{rec}} = 318$\,s, but different $d_{\text{ribbon}}$.  Respectively, from shortest to longest, the bundles are shown as solid, dotted, dashed, dot-dashed, and triple dot-dashed lines.  It is clear that the rising phase of each bundle is similar, but the cooling phase differs in each case in both \goes\ channels, where longer bundles take longer to decay.}
\label{fig:length_comparison}
\end{figure}

In Figure \ref{fig:length_comparison}, we compare five bundles of loops with equal reconnection time scale $\tau_{\text{rec}} = 318$\,s but variable $d_{\text{ribbon}}$.  The bundles consist of lengths $(2L)_{\text{min}} = [10, 15, 20, 25, 30]$\,Mm and respectively $(2L)_{\text{max}} = [20, 25, 30, 35, 40]$\,Mm, which corresponds to $d_{\text{ribbon}} = [9.5, 12.7, 15.9, 19.1, 22.3]$\,Mm, and we assume a slow ribbon speed $V_{\text{ribbon}} = 10$\,km\,s$^{-1}$ in order to emphasize the differences.  Respectively, from shortest to longest, the bundles are shown as solid, dotted, dashed, dot-dashed, and triple dot-dashed lines.

The differences between the individual bundles clearly show that while the rising phase of the light curves are similar, the time scales differ in both \goes\ channels.  The longer bundles take longer to decay in both channels (monotonically).  In other words, the lengths of the loops which comprise the bundle influence the values of the time scales, as one might expect from many previous loop cooling studies.  It is therefore clear that both the duration of reconnection and the lengths of the loops directly affect the \goes\ light curves. 

\subsection{The modeled correlation}
\label{subsec:correlation}

\begin{figure*}
\centering
\includegraphics[width=\linewidth]{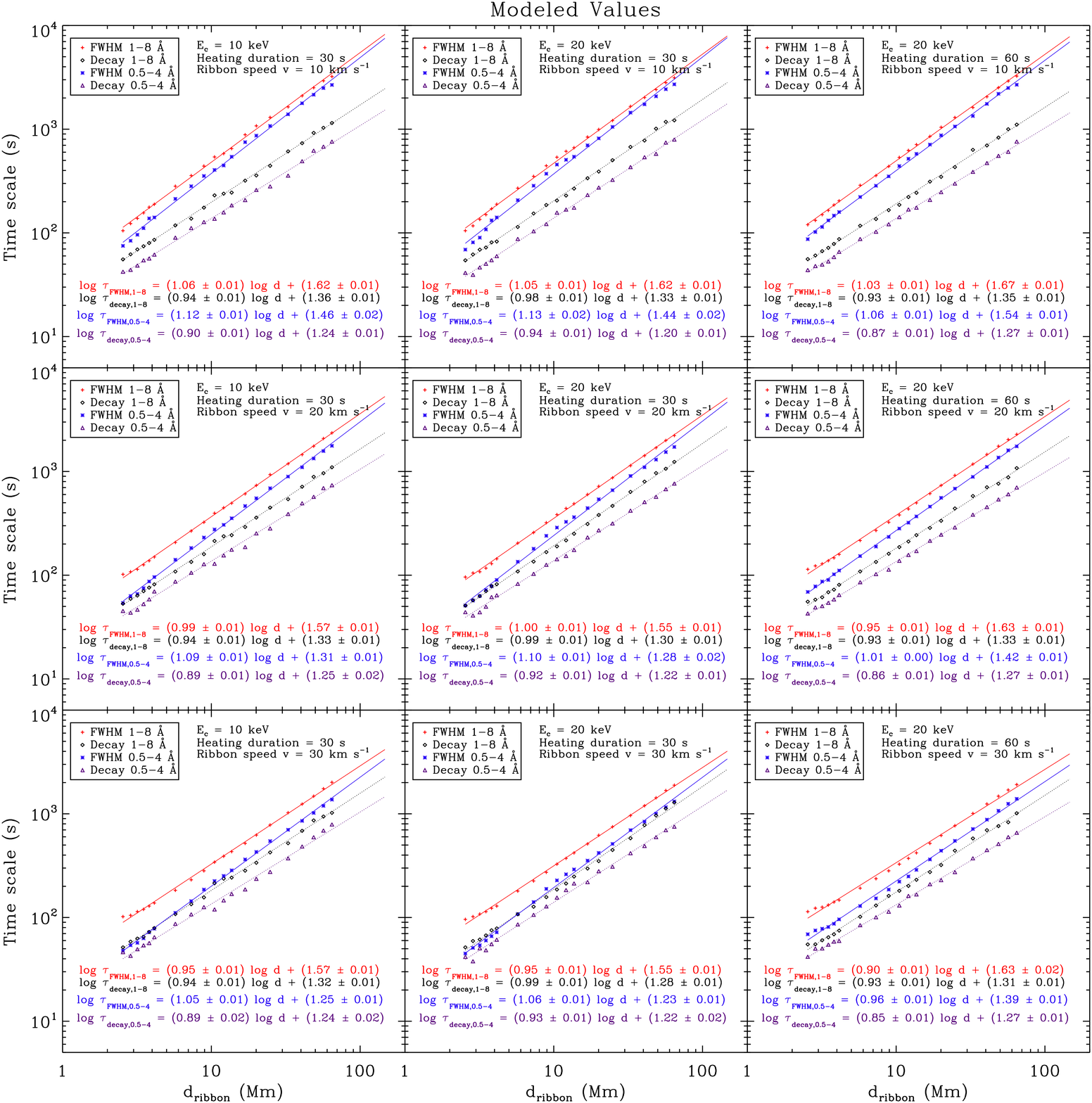}
\caption{9 scatter plots showing the FWHM and decay times as a function of separation between the ribbon centroids $d_{\text{ribbon}}$ as found from the numerical experiments, using a variety of parameters (see the text).  Compare these results against Figure \ref{fig:empirical}.}
\label{fig:scatter_plot}
\end{figure*}

In Figure \ref{fig:scatter_plot}, we summarize the results of many numerical experiments for different loop bundles with 9 scatter plots, showing the time scales as a function of separation of the ribbon centroids $d_{\text{ribbon}}$.  We include the data from both \goes\ channels.  The red plus signs (blue asterisks) show the FWHM $\tau_{\text{FWHM}}$ for the 1-8 (0.5-4) \AA\ channel, and the black diamonds (purple triangles) show the decay time $\tau_{\text{decay}}$.  The lines show the linear fit to each.  The top row synthesized the light curves with a ribbon speed $V_{\text{ribbon}} = 10$\,km\,s$^{-1}$, while the center and bottom used 20 and 30\,km\,s$^{-1}$.  The simulations in the first column used a cut-off energy $E_{c} = 10$\,keV, while the latter two columns used $20$\,keV.  Finally, the third column assumed a heating duration of $60$\,s, and the first two columns $30$\,s.

All 9 cases reproduce the approximately linear relation in all four time scales, though there is some variance.  For example, we find that there is reasonably good agreement between the observations and the case with $V_{\text{ribbon}} = 20$\,km\,s$^{-1}$, $E_{c} = 20$\,keV, and heating duration of $60$\,s.  For that case, the linear regression fits with 1-sigma uncertainties, also indicated on the plot, are:
\begin{align}
\log{\tau_{\text{FWHM, 1-8 \AA}}} &= (0.95 \pm 0.01) \log{d_{\text{ribbon}}} + (1.63 \pm 0.01) \nonumber \\
\log{\tau_{\text{decay, 1-8 \AA}}} &= (0.93 \pm 0.01) \log{d_{\text{ribbon}}} + (1.33 \pm 0.01) \nonumber \\
\log{\tau_{\text{FWHM, 0.5-4 \AA}}} &= (1.01 \pm 0.00) \log{d_{\text{ribbon}}} + (1.42 \pm 0.01) \nonumber \\
\log{\tau_{\text{decay, 0.5-4 \AA}}} &= (0.86 \pm 0.01) \log{d_{\text{ribbon}}} + (1.27 \pm 0.01) 
\end{align} 
\noindent In both channels, there is an approximately linear relation between both time scales and $d_{\text{ribbon}}$, with both the slopes and absolute values comparable to those measured from the observed light curves (Figure \ref{fig:empirical}).  $d_{\text{ribbon}}$ is directly related to the duration of the energy release due to reconnection $\tau_{\text{rec}}$ in this model (see Section \ref{sec:discussion}).  The separation between the centroids continues to grow as reconnection proceeds, which further implies that the time scale of reconnection $\tau_{\text{rec}}$ is also connected with the time scales of the \goes\ light curves.  To wit, the longer new loops are energized in a flare, the longer the SXR emission lasts.

There is little scatter in the calculated fits as compared to the observed trends, however.  We have made a number of simplifying assumptions that effectively reduce the scatter.  Though we have assumed it constant, the ribbon expansion speed generally varies with time \citep{asai2004}, with field strength \citep{xie2009}, and from flare to flare \citep{toriumi2017}, which would affect the rate at which loops are energized in the arcade model.  As each flare in the observational sample had different field strengths and occurred in different active regions of various sizes and types, it is clear that this assumption effectively reduces the variance in the calculated time scales.  Due to the different sizes of the active regions, our assumption that $(2L)_{\text{min}} = 3$\,Mm for all bundles is perhaps flawed, though it is not clear what value(s) would be more appropriate.    

We have also assumed that each loop has the same heating parameters, \textit{i.e.} that the electron beam parameters do not vary from loop to loop, or evolve with time.  We know from observational studies that the beam parameters do vary with time (\textit{e.g.} \citealt{holman2003,milligan2014}), but because HXR spatial resolution is limited, it is not clear how they change amongst individual loops.  How the parameters vary from loop to loop is an on-going area of research that is at present poorly understood, because of this difficulty of resolving individual loops in a flare, even with high-resolution extreme ultraviolet satellites \citep{warren2006,warren2016,reep2016b}.  Another potential issue is that we have assumed only one source of heating of the plasma, whereas there is good reason to suspect that other mechanisms contribute to energy transport, including shock heating \citep{longcope2009,longcope2016}, magnetic wave damping \citep{reeprussell2016,kerr2016}, and \textit{in situ} heating driving thermal conduction fronts \citep{longcope2014}.

\section{Discussion}
\label{sec:discussion}

A clear linear correlation has been found between the separation of ribbon centroids and the FWHM and e-folding decay times of \goes\ light curves, which has been tested for large (above M5) flares.  It was not clear how the correlation arises, or what mechanisms are involved.  Flare models consisting of a single loop cool significantly faster than the observed trends, and therefore we developed an expanding arcade model to examine them.  The model reproduces the linear trend: the FWHM and decay times are longer for larger centroid separations $d_{\text{ribbon}}$.  $d_{\text{ribbon}}$ is determined primarily by the duration of reconnection, while the rate at which individual loops cool is determined primarily by their lengths (\textit{e.g.} \citealt{cargill1995}).

As the reconnection event proceeds, forming new loops and causing the expansion of the arcade, the ribbons of the flare continue to separate.  For a constant ribbon expansion speed, the duration of reconnection determines the centroid separation $d_{\text{ribbon}}$, and therefore is connected directly to the time scales.  In general, though, the ribbon expansion speed is not constant, which is one cause of scatter in the observed trends.  
	
Coronal loops cool primarily through three processes: thermal conduction ($\tau_{C} \propto L^{2}$), radiation ($\tau_{R}$ independent of length), and a draining enthalpy flux ($\tau_{V} \propto L$), each successively dominating the cooling of an individual loop \citep{bradshaw2010}, which causes the total cooling time to depend strongly on length ($\propto L^{5/6}$, \citealt{cargill1995}).  In a flare bundle with many loops, however, it is unclear which of these dominates the apparent cooling in light curves.  We have found numerically that the FWHM and decay times increase for bundles of longer loops, even when the reconnection time scale remains constant.  This demonstrates that both the reconnection time scale and the cooling of the individual loops determine the total time scales, and therefore we conclude that both cause the observed linear relation.

There are still a few open questions.  \citet{sainthilaire2010} found a weak quadratic correlation between the HXR burst duration and the HXR foot-point separation in a sample of 53 large flares seen with \rhessi: $\tau_{\text{HXR}} \propto d_{\text{HXR}}^{2}$.  Under the presumption that the HXR separation $d_{\text{HXR}}$ is approximately the same as $d_{\text{ribbon}}$, then we would have $\tau_{\text{HXR}} \propto \tau^{2}$ (where $\tau$ could be either time scale).  This is a relation that should be carefully checked in the future, as there would be important implications regarding heating durations, the number of strands in a flare, and the reconnection process itself.

We plan a more extensive parameter survey, where we allow many other beam parameters to vary from loop to loop, as well as a time-varying ribbon expansion speed, as each parameter has an effect on the \goes\ light curve (\textit{e.g.} \citealt{reep2013}).  We also will examine two-phase heating, where it appears that all loops comprising the flare are heated both impulsively and by a gradual and significantly weaker heat source (\textit{e.g.} \citealt{qiu2016}).  

To more accurately model a given flare, it is important to determine the loop lengths, as demonstrated in Section \ref{subsec:light_curves}.  We therefore would like to incorporate a method to obtain the loop length distribution more directly, rather than making assumptions.  In practice, this could be accomplished with a loop tracing algorithm \citep{aschwanden2010}, non-linear force-free field modeling \citep{wiegelmann2012}, or flaring active region modeling \citep{toriumi2017b}.  It would also be an important test to confirm observationally that $d_{\text{ribbon}}$, the distance between the centroids of the two ribbons, is representative of the distribution of loop lengths within a flare.


\acknowledgments The authors would like to thank Nick Crump for assistance in the data processing, Harry Warren for comments on an early version of this paper, and the anonymous referee for many comments that have broadened and improved the scope of this work.  This research was performed while JWR held an NRC Research Associateship award at the US Naval Research Laboratory with the support of NASA.  This work was supported by JSPS KAKENHI Grant Numbers JP16K17671, JP15H05814.  CHIANTI is a collaborative project involving George Mason University, the University of Michigan (USA) and the University of Cambridge (UK).  This research has made use of NASA's Astrophysics Data System.


\bibliographystyle{apj}

\end{document}